\documentclass[aps,prl,twocolumn,superscriptaddress,floatfix]{revtex4-2}

\usepackage{amsmath}
\usepackage{amssymb}
\usepackage{graphicx}
\usepackage{bm}
\usepackage{xcolor}
\usepackage{svg}
\usepackage{float}
\usepackage{subcaption}
\usepackage[colorlinks=true,linkcolor=blue,citecolor=blue,urlcolor=blue]{hyperref}

\newcommand{\G}{\Gamma}
\newcommand{\Gr}{\Gamma_r}

\begin{document}

\title{Collective phases in overdamped magnetic self-propelled spherocylinders}

\author{Francisca Guzm\'an-Lastra}
\affiliation{Departamento de F\'isica, Facultad de Ciencias, Universidad de Chile, Santiago, Chile.}

\author{N\'estor Sep\'ulveda}
\affiliation{School of Engineering and Sciences, Universidad Adolfo Ib{\'a}{\~n}ez, Diagonal las Torres 2640, Pe\~{n}alolén, Santiago, Chile.}

\date{\today}

\begin{abstract}
We study the collective dynamics of self-propelled spherocylinders carrying 
magnetic dipole moments in two dimensions. Magnetic interactions are modeled 
as two opposite monopoles $\pm Q$ separated by a distance $\ell$ along the 
particle director, a dumbbell model that remains well-defined at short range 
and introduces an explicit geometric lever arm for the magnetic torque. This 
approach, combined with the elongated particle geometry, produces a torque 
that competes with steric alignment in a manner inaccessible to point-dipole 
or disk models. By independently varying monopole separation and dipole 
strength (parameters that map directly onto the geometry and magnetization 
of cylindrical magnets) we show that the system navigates a rich landscape 
of collective states: gas, polar flock, chain, vortex-alignment, and 
locked-dimer phases. Our results establish that particle elongation and 
distributed magnetic charge together provide a minimal, experimentally 
accessible set of tuning knobs for controlling coherent states in 
magnetic active matter, with direct implications for the design of 
self-organized magnetic microswimmers and active colloidal assemblies.
\end{abstract}

\maketitle


Active matter systems display a rich variety of collective behaviors with no 
equilibrium counterpart~\cite{ramaswamy2010, gompper2025}. Polar flocking, 
rotating-cluster formation, and mesoscale turbulence all arise from local 
interactions among motile particles, yet the symmetry of the resulting 
macroscopic state depends sensitively on which interactions are 
present~\cite{vicsek1995, marchetti2013, ramaswamy2010}.

Particle shape alone can drive collective order. Self-propelled rods interact 
through steric excluded-volume collisions that align nearby directors, producing 
a transition from a disordered gas to a polar flocking state whose threshold 
depends on aspect ratio~\cite{peruani2006, weitz2015self}. At large system 
sizes the polar phase acquires substructure including giant clusters and 
propagating density bands~\cite{barberis2016, broker2024collective}, and at 
high packing fraction the dynamics become turbulent~\cite{wensink2012, 
alert2022active, zantop2022emergent}. Experiments on light-driven self-propelled 
rods confirm that shape anisotropy alone governs transitions among swarming, 
turbulence, flocking, and jamming~\cite{shelke2026shape}, establishing aspect 
ratio as a primary tuning parameter for collective order.

Magnetic interactions, which arise naturally in magnetotactic bacteria and can 
be engineered into colloidal microswimmers, magnetic nanoparticles, and 
macroscopic granular robots, add a further layer of control~\cite{vincenti2019magnetotactic, waisbord2016destabilization,petroff2022phases,jin2021collective, mandal2018magnetic}. Dipole-dipole 
interactions are inherently anisotropic: head-to-tail coupling drives chain 
formation and polar alignment, while side-by-side anti-parallel configurations 
select vortex order~\cite{kokot2017, kaiser2017, kokot2015emergence}. Which 
phase emerges is governed by the ratio of dipole strength to self-propulsion 
activity~\cite{liao2020dynamical, vanesse2023collective}, the dipole orientation 
relative to the propulsion axis~\cite{vanesse2023collective, martinez2018emergent, 
tierno2014recent}, particle density~\cite{liao2020dynamical, han2020}, 
confinement geometry~\cite{telezki2020simulations}, and self-rotation 
frequency~\cite{liao2021emergent}, yielding a sequence of gas, cluster, chain, 
flocking, and vortex states. Non-equilibrium fission and fusion of magnetic 
microswimmer clusters reveal additional scenarios beyond this 
sequence~\cite{guzman2016fission, martinez2018emergent, tierno2021transport, 
klumpp2019swimming}. These phases have been observed experimentally in rolling 
magnetic colloid dimers~\cite{kokot2017}, flocking ferromagnetic 
colloids~\cite{kaiser2017, kokot2015emergence}, and multi-vortex assemblies of 
active rollers~\cite{han2020}. Applying an external field enriches pattern formation but introduces an additional experimental constraint~\cite{koessel2020emergent, 
koessel2019controlling, telezki2025patterns, parage2025modulation}.
At the macroscopic scale, hexbugs equipped with 
elongated neodymium magnets provide an experimentally accessible realization: 
both disk-shaped and elongated variants self-organize into chains, rings, and 
vortex states~\cite{obreque2024dynamics, musacchio2026fluidization, 
sepulveda2021bioinspired}, with particle elongation favoring rotating over polar 
order, an experimental signature relevant in dilute regimes.

The point-dipole approximation diverges at short separations and eliminates 
the geometric torque that arises when the magnetic charge distribution is 
spatially extended relative to the particle body~\cite{sepulveda2021bioinspired}. 
The dumbbell representation (two opposite monopoles $\pm Q$ separated by 
$\ell$ along the particle axis) was introduced in spin-ice physics to describe 
emergent magnetic monopoles~\cite{castelnovo2008magnetic} and has since been 
applied to macroscopic frustrated magnets~\cite{mellado2012macroscopic}, 
anisotropic magnetic colloids~\cite{kantorovich2013influence}, and dipolar 
chains with hysteretic response~\cite{concha2018designing}. In elongated 
particles, $\ell$ sets an independent lever arm for the magnetic torque absent 
in the point-dipole limit. By independently tuning $\ell$ and $Q$ (parameters 
that map directly onto the elongation and magnetization of cylindrical 
magnets) one can in principle navigate the full landscape of collective phases 
without varying density or activity. Whether this control is achievable, and 
what new phases it might unlock, remains unexplored in active systems.

Here we simulate $N=500$ overdamped self-propelled spherocylinders with a
dumbbell magnetic dipole, scanning an $8\times10$ grid in monopole separation
$\Pi_1$ and dimensionless dipole strength $\Pi_2$ (details in the Supplemental Material~\cite{SM}). We identify five
collective phases and derive closed-form phase boundaries from the magnetic
pair contact energies, providing a unified energetic picture of shape-driven
and torque-driven order. Our results unify phases reported across disparate
active magnetic matter
experiments~\cite{kaiser2017,han2020,liao2020dynamical,vanesse2023collective,telezki2020simulations}
and establish design principles for programmable magnetic active
matter~\cite{tierno2021transport,gao2025soft,pramanik2024nature}.

\begin{figure}[ht!]
  \centering
  \includegraphics[width=0.85\columnwidth]{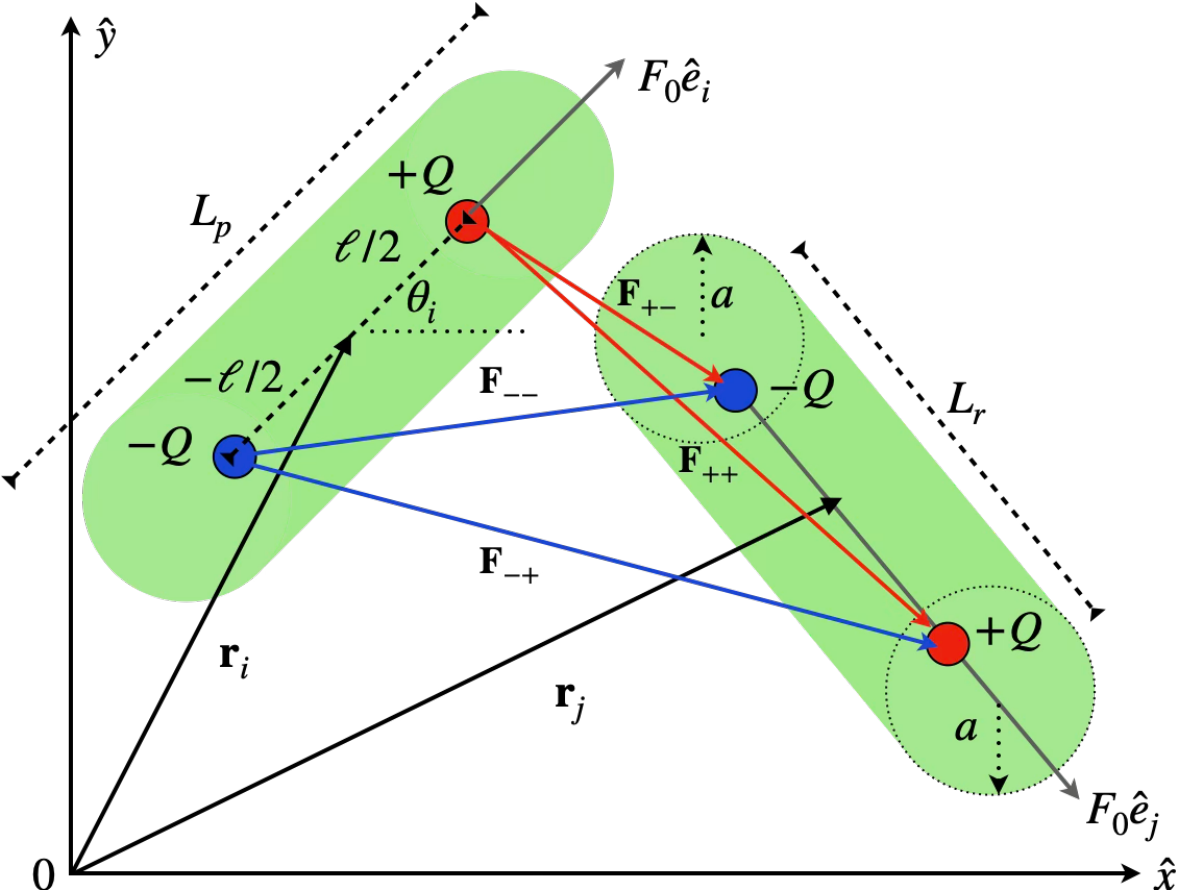}
  \caption{Two interacting self-propelled spherocylinders ($i$, $j$) carrying
    magnetic dipole moments modeled as two opposite monopoles $\pm Q$ (positive
    in red, negative in blue) separated by a distance $\ell$ along the particle
    director.}
  \label{fig1}
\end{figure}

{\bf\emph{Model:}} Each particle $i$ ($i=1,\ldots,N$) is a two-dimensional spherocylinder
(Fig.~\ref{fig1}) with rod backbone length $L_r$, hemisphere radius $a$,
total length $L_p=L_r+2a$, and orientation $\hat{\mathbf{e}}_i = \bigl(\cos\theta_i,\,\sin\theta_i\bigr)$ where $\theta_i$ is the angle with the $x$-axis. 
Particles are initially uniformly distributed in position and orientation with packing fraction $\phi\approx 0.18$. Simulations are carried out with periodic boundary conditions; see the Supplemental Material~\cite{SM} for details.
The particles are self-propelled with active force $F_0$ acting along their body axis $\hat{\mathbf{e}}_i$. The overdamped equations of motion are, 
\begin{align}
  \Gamma\,\dot{\mathbf{r}}_i &= F_0\hat{\mathbf{e}}_i
    + \mathbf{F}_i^{\rm rep} + \mathbf{F}_i^{\rm mag},
  \label{eq:eom_r}\\
  \Gamma_r\,\dot\theta_i &= \tau_i^{\rm rep} + \tau_i^{\rm mag},
  \label{eq:eom_theta}
\end{align}
with translational friction $\Gamma$ and rotational friction $\Gamma_r$~\cite{liao2020dynamical}.
When the minimum surface-to-surface distance $d_{ij}$ between the
backbone segments of rods $i$ and $j$ is smaller than the contact
diameter $2a$, a linear spring repulsion acts perpendicular to
the surface $\mathbf{F}_{i\leftarrow j}^{\rm rep} =
    -E_{\rm int}\bigl(2a - d_{ij}\bigr)\,\hat{\mathbf{n}}_{ij}$, where $E_{\rm int}$ is the steric stiffness. The repulsive torque associated with this interaction is 
$\tau_{i\leftarrow j}^{\rm rep}=\boldsymbol{\rho}_i^{\rm cp}
\times\mathbf{F}_{i\leftarrow j}^{\rm rep}$,
where $\boldsymbol{\rho}_i^{\rm cp}$ is the vector from the center of mass of particle $i$ to the contact point with particle $j$.

Each particle carries a magnetic moment modeled by two opposite monopoles 
$\pm Q$ at positions $\mathbf{r}_i \pm (\ell/2)\hat{\mathbf{e}}_i$ (Fig.~\ref{fig1}), yielding 
a net dipole $\mathbf{m}_i = Q\ell\,\hat{\mathbf{e}}_i$ aligned with the 
propulsion axis~\cite{kaiser2015active, guzman2016fission, liao2020dynamical, 
vanesse2023collective}. Unlike point-dipole models, this dumbbell 
representation~\cite{castelnovo2008magnetic, mellado2012macroscopic, 
concha2018designing} regularizes the short-range divergence and introduces 
an explicit geometric lever arm $\ell$ for the magnetic torque. The 
interaction between monopole $\alpha$ on particle $i$ and monopole $\beta$ 
on particle $j$ is
\begin{equation}
  \mathbf{F}_{\alpha\beta} =
    \frac{\mu_0}{4\pi}\,
    \frac{q_{i\alpha}\,q_{j\beta}}
         {|\mathbf{r}_{i\alpha} - \mathbf{r}_{j\beta}|^3}\,
    \bigl(\mathbf{r}_{i\alpha} - \mathbf{r}_{j\beta}\bigr),
  \label{eq:monopole}
\end{equation}
summed over all four sign combinations. The total magnetic force and torque 
on particle $i$ due to particle $j$ are
\begin{align}
  \mathbf{F}_{i\leftarrow j}^{\rm mag}
    &= \mathbf{F}_{++} + \mathbf{F}_{+-}
     + \mathbf{F}_{-+} + \mathbf{F}_{--},
  \label{eq:mag_force}\\
  \tau_{i\leftarrow j}^{\rm mag}
    &= \boldsymbol{\rho}_{i+}
       \times\bigl(\mathbf{F}_{++}+\mathbf{F}_{+-}\bigr)
     + \boldsymbol{\rho}_{i-}
       \times\bigl(\mathbf{F}_{-+}+\mathbf{F}_{--}\bigr),
  \label{eq:mag_torque}
\end{align}
where $\boldsymbol{\rho}_{i\pm} = \pm(\ell/2)\hat{\mathbf{e}}_i$. In the 
far-field limit $r_{ij} \gg \ell$ the model recovers standard dipole--dipole 
interactions.

Three dimensionless groups control the collective behavior:
\begin{equation}
  \Pi_1 = \frac{\ell}{L_r},
  \quad
  \Pi_2 = \frac{\mu_0\,Q^2\,\G}{4\pi\,F_0\,\Gr},
  \quad
  \Pi_3 = \frac{E_{\rm int}L_r}{F_0}.
  \label{eq:pi12}
\end{equation}
$\Pi_1$ is the ratio of charge separation to rod length; $\Pi_2$ compares 
the magnetic and self-propulsion angular velocities, with $\Pi_2 \gg 1$ 
indicating magnetically dominated orientational dynamics; and $\Pi_3$ 
compares steric restoring force to self-propulsion, setting the contact time 
$\tau_{\rm con} = \G/E_{\rm int}$ during which the magnetic torque builds 
rotational coherence. We fix $\Pi_3 = 150$, which corresponds to the largest 
contact time for which coherent vortex alignment patterns remain stable. 
The influence of $\Pi_3$ is examined systematically in the Supplemental Material~\cite{SM}. A complete summary of all model parameters is provided 
in Table~S1 of the Supplemental Material.
\begin{figure*}[t]
  \centering
\includegraphics[width=\textwidth]{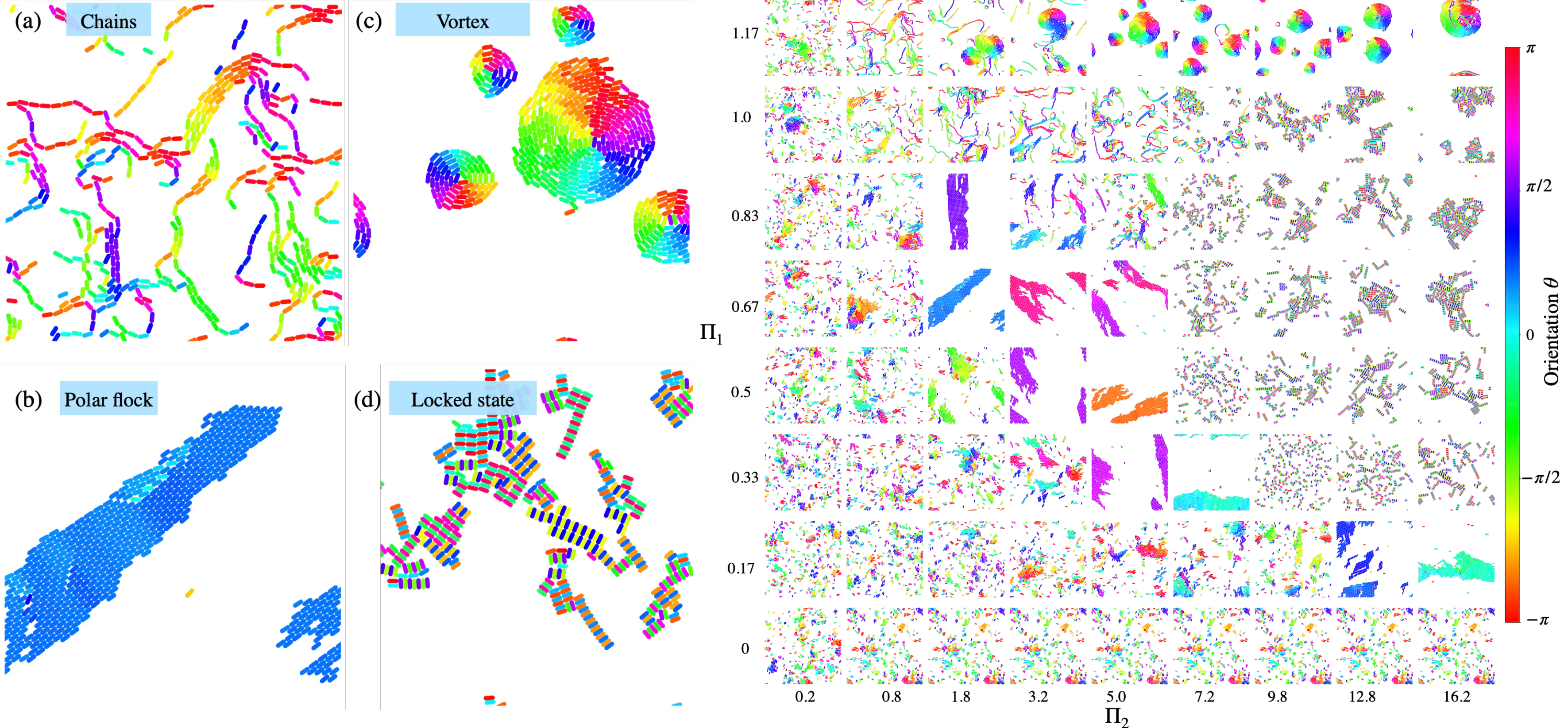}
  \caption{Steady-state configurations at $\Pi_3=150$; particles are colored by
orientation angle $\theta\in(-\pi,\pi]$ using a cyclic HSV colormap
(shared colorbar). A nearly uniform hue signals polar alignment, swirling 
opposing hues within a cluster signal vortex rotation, and random hues 
indicate disorder. \textbf{Left:} Representative snapshots of the four 
collective states. \emph{Chain} ($\Pi_2\approx0.8$, $\Pi_1\approx1.17$): 
elongated quasi-linear aggregates driven by head-to-tail attraction, without 
vortex rotation or global polar alignment. \emph{Coherent vortex-alignment} 
($\Pi_2\approx5.0$, $\Pi_1\approx1.17$): intra-cluster vortex rotation. 
\emph{Polar flocking} ($\Pi_2\approx1.8$, $\Pi_1\approx0.67$): global 
director alignment. \emph{Locked} ($\Pi_2\approx12.8$, $\Pi_1\approx0.83$): 
dense compact aggregates formed by antiparallel binding suppress all 
collective motion. \textbf{Right:} Full $\Pi_2\times\Pi_1$ phase diagram for 
$E_{\rm int}=100$; columns increase in $\Pi_2$ (left to right) and rows 
increase in $\Pi_1$ (bottom to top). The $\Pi_1=0$ row remains disordered 
at all $\Pi_2$ because $\ell=0$ suppresses the magnetic interaction.}
  \label{fig2}
\end{figure*}

\begin{figure}[ht!]
  \centering
  \includegraphics[width=0.8\columnwidth]{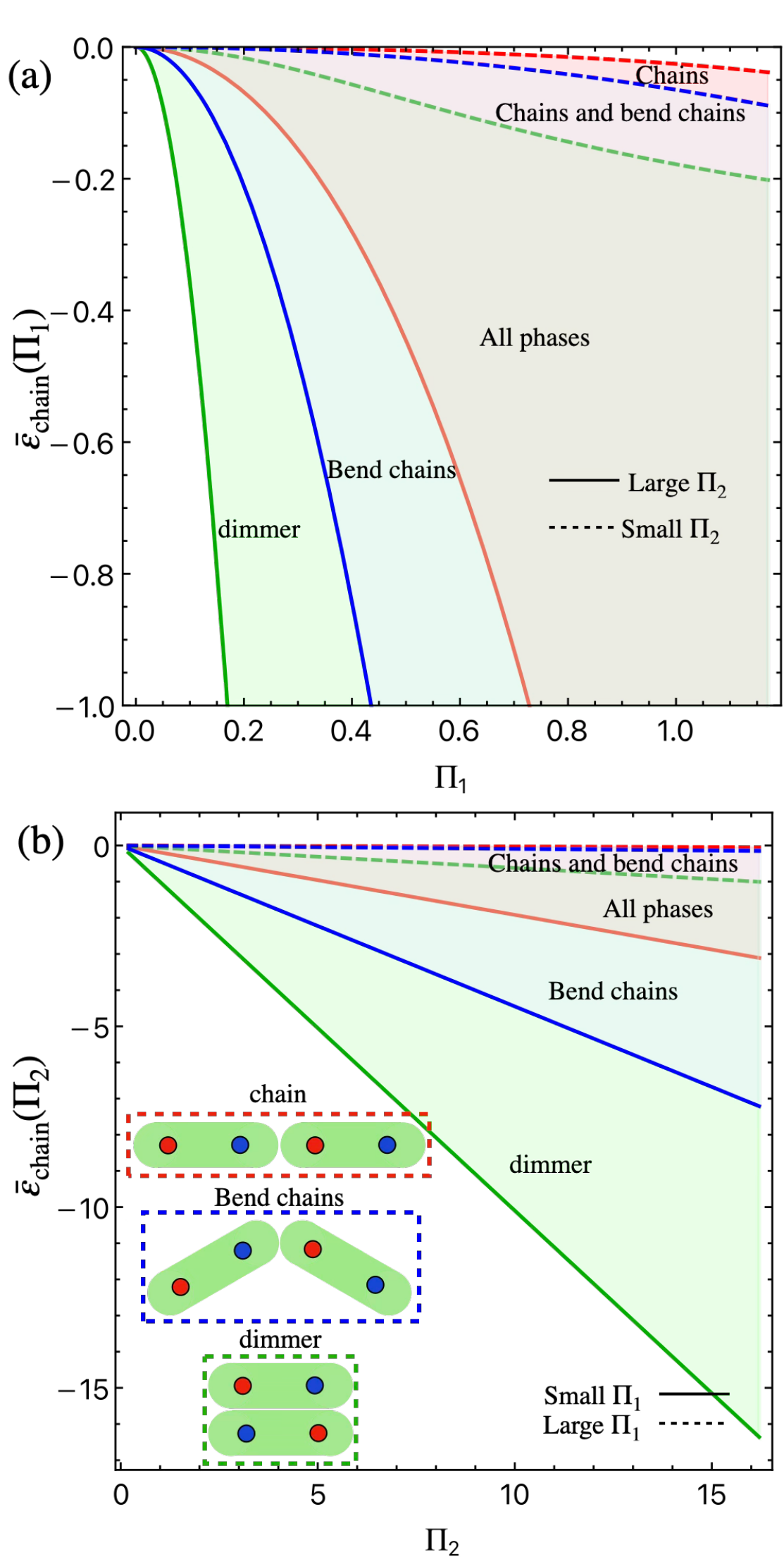}
  \caption{Dimensionless two-particle contact energies $\bar\varepsilon_{\rm chain}$
(red), $\bar\varepsilon_{\rm bend}$ (blue), and $\bar\varepsilon_{\rm
blocked}$ (green), normalized by $F_0 L_r$
(Eqs.~\eqref{eq:ebar_chain}--\eqref{eq:ebar_blocked}), as a function of
(a)~$\Pi_1$ and (b)~$\Pi_2$. In each panel, solid and dashed lines
correspond to the limiting large and small values of the other parameter,
respectively. Shaded regions in (a) indicate where each configuration is
energetically accessible: chain only (pink), bent chain only (light blue), and dimer or blocked pair (light green). Insets in (b) illustrate the three contact geometries; red and
blue circles denote positive and negative magnetic charges.}
  \label{fig3}
\end{figure}

\begin{figure}[t]
  \centering
  \includegraphics[width=\columnwidth]{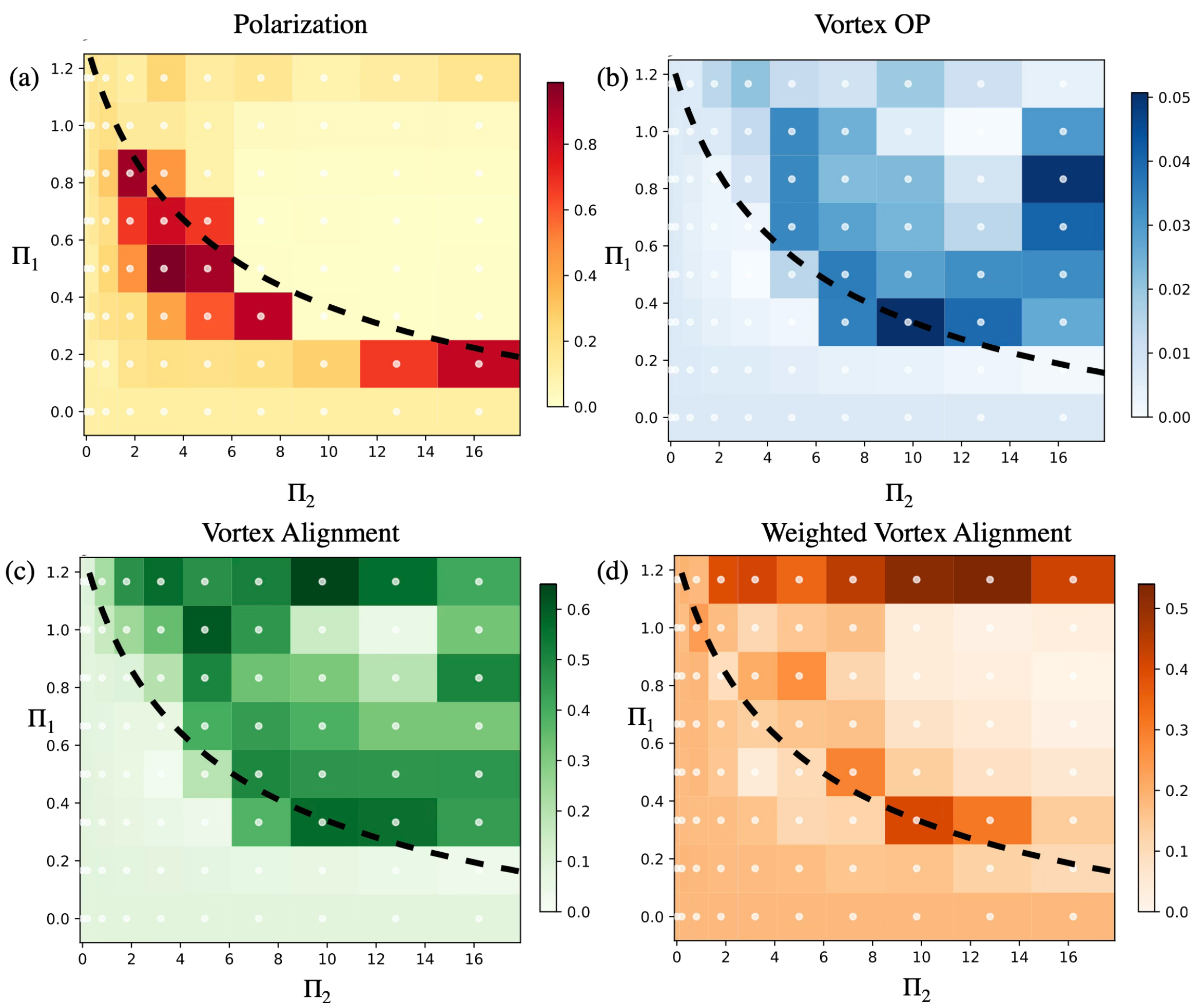}
  \caption{Phase diagram in the $\Pi_2\times\Pi_1$ plane for $\Pi_3=150$. The four 
panels show the scalar order parameters: (a)~polarization $P$; (b)~vortex 
order parameter $\Phi_V$; (c)~vortex alignment $\Phi_A$; (d)~weighted vortex 
alignment $\Phi_A^w$. White dots mark the 80 validated state points 
($\Pi_2\in\{0,0.2,\ldots,16.2\}$, $\Pi_1\in\{0,0.17,\ldots,1.17\}$; state 
points with $\Pi_1\ge 1.33$ are excluded; see the validity analysis
  in the Supplemental Material \cite{SM}). The black dashed line 
marks the contour $|\bar\varepsilon_{\rm chain}|=1$ (see
  Fig.~\ref{fig3}).}
  \label{fig4}
\end{figure}

\begin{figure}[t]
  \centering
  \includegraphics[width=\columnwidth]{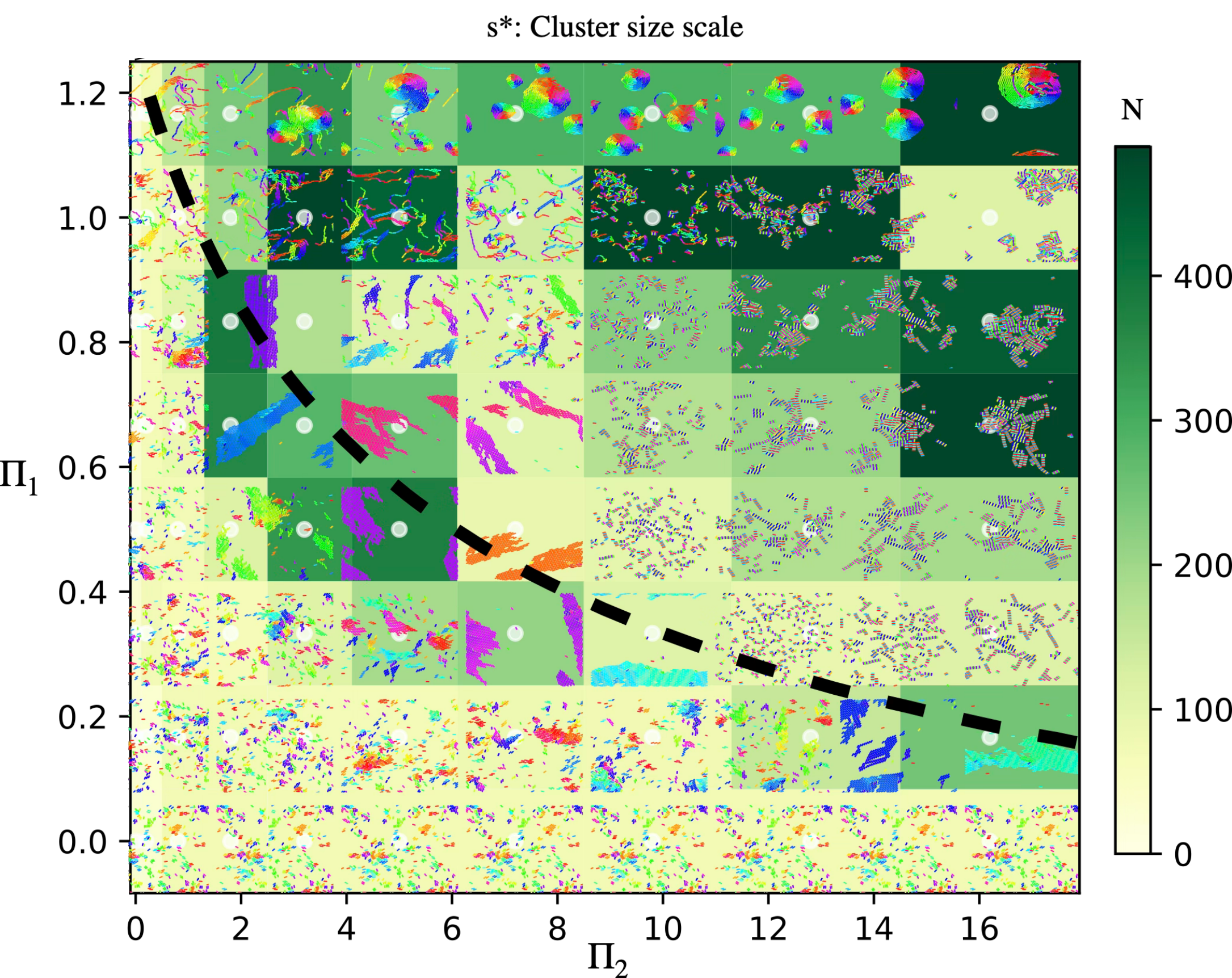}
  \caption{Cluster size scale $s^*$ in the $\Pi_2\times\Pi_1$ plane for $\Pi_3=150$,
with simulation snapshots superimposed at each of the 80 state points.
$s^*$ is the mean size of the largest cluster normalized by $N=500$,
ranging from disordered (light, $s^*\approx0$) to fully aggregated
(dark, $s^*\approx N$). Labels indicate the observed collective regimes.
The black dashed line marks the contour $|\bar\varepsilon_{\rm chain}|=1$ (see
  Fig.~\ref{fig3}); large-aggregate states
($s^*\gg1$) lie predominantly above it, with gas and gas-and-cluster
phases below.}
  \label{fig5}
\end{figure}

{\bf \emph{Results and Discussion:}} Figure~\ref{fig2} maps the collective behavior across the $(\Pi_1,\Pi_2)$ 
plane at $\Pi_3=150$. Five states emerge as $\Pi_1$ and $\Pi_2$ increase: 
a disordered gas (Movie~S1), 
a chain state of head-to-tail aggregates (Movie~S2), 
a polar flocking state (Movie~S3), a coherent vortex-alignment 
state of co-rotating clusters (Movie~S4), and a quasi-arrested locked state 
in which strongly bound aggregates suppress but do not fully extinguish 
translational motion (Movie~S5). Polar order and vortex alignment occupy 
complementary regions, reflecting the competition between head-to-tail 
attraction and the antiparallel side-by-side torque. Quantitative phase 
diagrams and corresponding order parameter maps are presented in Figs.~(\ref{fig4},\ref{fig5}). 
Four values of $\Pi_3$ are compared in the Supplemental Material~\cite{SM}.

The phase boundaries can be rationalized through the magnetic pair interaction 
energies at three characteristic contact geometries: head-to-tail chain, bent 
chain, and antiparallel blocked pair (Fig.~\ref{fig3}; detailed derivations, together with the explicit expression for $\bar\varepsilon_{\rm bend}$, are provided in the Supplemental Material~\cite{SM}). Normalizing by the propulsion work $F_0 L_r$, 
the chain and antiparallel blocked pair energies are

\begin{equation}
  \bar\varepsilon_{\rm chain}
  = -\frac{6\Pi_1^2\,\Pi_2}{5(25-9\Pi_1^2)},
  \label{eq:ebar_chain}
\end{equation}

\begin{equation}
  \bar\varepsilon_{\rm blocked}
  = -\frac{2\Pi_2}{3}
    \!\left(\frac{1}{2}-\frac{1}{\sqrt{9\Pi_1^2+4}}\right).
  \label{eq:ebar_blocked}
\end{equation}

Both vanish at $\Pi_1=0$ and are linear in 
$\Pi_2$ (Fig.~\ref{fig3}). The ratio

\begin{equation}
  R(\Pi_1) \equiv
  \frac{|\bar\varepsilon_{\rm blocked}|}{|\bar\varepsilon_{\rm chain}|}
  = \frac{5(25-9\Pi_1^2)}{9\Pi_1^2}
    \!\left(\frac{1}{2}-\frac{1}{\sqrt{9\Pi_1^2+4}}\right)
  \label{eq:ratio}
\end{equation}

is \emph{independent of $\Pi_2$} and decreases monotonically from 
$R\approx 7.4$ at $\Pi_1=0.17$ to $R\approx 1.3$ at $\Pi_1=1.17$. 
This independence is the key to reading the phase diagram: the energetic 
advantage of the antiparallel dimer over the chain bond is set entirely 
by $\Pi_1$, which is why the polar-to-vortex crossover appears as a 
near-horizontal boundary in Fig.~\ref{fig2}.

\emph{Gas phase and gas-and-cluster phase:} This regime is characterized by a predominantly disordered gas
coexisting with small clusters that exhibit local polar alignment
(Fig.~\ref{fig2}). Two limiting cases immediately clarify its
boundaries. When $\Pi_1 = 0$, corresponding to the point-dipole
approximation with vanishing magnetic moment, the system reduces to a
collection of active particles interacting exclusively through steric
forces, regardless of how large the magnetic charge becomes with
increasing $\Pi_2$. Analogously, when $\Pi_2 = 0$ the particle
charges vanish identically, so that even as $\ell$ grows the particles
remain magnetically non-interacting. In both limits
$|\bar\varepsilon_{\rm chain}|$ and $|\bar\varepsilon_{\rm blocked}|$
vanish identically (Eqs.~\eqref{eq:ebar_chain}--\eqref{eq:ebar_blocked}):
no contact geometry is capable of trapping particles and the system
remains disordered regardless of $\Pi_2$.

Within the region $\Pi_2 < 1$, where magnetic interactions have not
yet come to dominate over steric-induced reorientation, increasing
$\Pi_1$ drives the system progressively into the gas-and-cluster
regime: as $\ell$ grows, excluded-volume alignment among the
spherocylinders becomes increasingly effective, promoting transient
local order without generating global coherence~\cite{liao2020dynamical, vanesse2023collective} (see Movie~S1). Two transitions
depart from this regime: for small $\ell$, increasing $\Pi_2$ drives
the system toward a polar-flock state, whereas for $\Pi_2 < 1$ and
$\Pi_1 \gtrsim 1$ the growing magnetic interaction strength instead
drives a transition to the chain state. The scalar order parameters remain 
close to their disordered values, reflecting the absence of global organization 
despite frequent local alignment events. Accordingly, the polarization, vortex 
order parameter, and vortex alignment parameter remain near zero in this phase 
(Fig.~\ref{fig4}), as does the characteristic cluster size (Fig.~\ref{fig5}). 
As shown in Fig.~\ref{fig5}, this region lies entirely below the black dashed 
curve marking the $|\bar\varepsilon_{\rm chain}|=1$ threshold (Eq.~\eqref{eq:ebar_chain}), 
which corresponds to one of the metastable configurations previously identified in 
magnetic-disk systems with a point dipole~\cite{liao2020dynamical,liao2021emergent,vanesse2023collective}. 
The chain and polar-flock states appear near or above this threshold, as discussed below.

\emph{Chain state.} For $\Pi_1 \gtrsim 1$ and low-to-moderate $\Pi_2$, 
head-to-tail attraction drives particles into elongated quasi-one-dimensional 
aggregates (Fig.~\ref{fig2}, Movie~S2)~\cite{liao2020dynamical,telezki2020simulations}. 
At large $\Pi_1$ the magnetic charges sit near the spherocylinder tips, 
bringing opposite charges on neighboring particles into close proximity in 
the head-to-tail configuration (Fig.~\ref{fig3}(a)), and $R(\Pi_1)$ 
(Eq.~\eqref{eq:ratio}) remains large enough that the antiparallel torque 
cannot yet generate vortex patterns. Different chains propel in independent 
directions, so $P$ and the vortex order parameters $\Phi_V$, $\Phi_A$ are 
all suppressed (Fig.~\ref{fig4}); the slight residual polarization reflects 
individual chains translating coherently before reorienting~\cite{liao2020dynamical,vanesse2023collective,telezki2020simulations}.

\emph{Polar flocking state.} For $\Pi_1 \lesssim 1$ and $\Pi_2 \gtrsim 1$, 
the magnetic charges remain close to the particle backbone, weakening the deep 
head-to-tail well that would lock particles into persistent chains. With 
$|\bar\varepsilon_{\rm blocked}|<1$ throughout this ridge, the antiparallel 
binding energy has not yet overcome self-propulsion, so head-to-tail coupling 
instead promotes global director alignment (Fig.~\ref{fig2}, 
Movie~S3)~\cite{kaiser2017,liao2020dynamical}. A system-spanning cluster forms 
and propagates coherently, reflected in high $P$ and large $s^*$ 
(Figs.~\ref{fig4},\ref{fig5}), while $\Phi_V$ and $\Phi_A$ remain small 
($\Phi_A \approx 0.06$--$0.16$), confirming purely translational collective 
motion. The upper boundary of this phase is set by $|\bar\varepsilon_{\rm
blocked}|=1$: once antiparallel binding exceeds
self-propulsion the coherent flock dissolves into arrested aggregates.

\emph{Coherent vortex-alignment state.} At $\Pi_1 \gtrsim 1$ and $\Pi_2 \sim 
\mathcal{O}(10)$, the magnetic torque strongly dominates active reorientation 
and $R(\Pi_1)$ has decreased sufficiently that the antiparallel side-by-side 
torque wins during collisions, tipping the balance from polar to vortex order 
(Fig.~\ref{fig2}, Movie~S4). Because both contact energies scale linearly with 
$\Pi_2$, this crossover is governed primarily by $\Pi_1$, explaining the 
near-horizontal polar-to-vortex boundary in Fig.~\ref{fig2}. The antiparallel 
dimer (Fig.~\ref{fig3}(b)), metastable for $\Pi_1 \gtrsim 0.2$, serves as 
the nucleation unit from which larger rotating clusters grow. Each cluster 
rotates coherently, but opposite handedness across clusters averages $P$ to 
zero; simultaneously $\Phi_V$ and $\Phi_A$ peak ($\Phi_A \approx 0.55$--$0.56$, 
Fig.~\ref{fig4}) and $s^*$ is large (Fig.~\ref{fig5}). With 
$|\bar\varepsilon_{\rm blocked}|<1$ throughout, the antiparallel binding 
assists rotational coherence without permanently arresting translation. As 
shown in the Supplemental Material~\cite{SM}, increasing $\Pi_3$ progressively 
destroys this phase, confirming that collision dynamics (not the equilibrium 
energy landscape alone) govern its stability.

\emph{Locked state.} For $\Pi_2 \gtrsim 10$ the condition 
$|\bar\varepsilon_{\rm blocked}|=1$ is crossed and the system enters a locked 
state (Fig.~\ref{fig2}, Movie~S5), with onset well predicted by 
\begin{equation}
  \Pi_2^{\rm lock}(\Pi_1)
  = \frac{3\sqrt{9\Pi_1^2+4}}{\sqrt{9\Pi_1^2+4}-2}.
  \label{eq:pi2lock}
\end{equation}
This threshold is monotonically decreasing in $\Pi_1$: the locked state is 
absent for $\Pi_1 \lesssim 0.3$ within the scanned range and falls to 
$\Pi_2^{\rm lock} \approx 5.9$ at $\Pi_1=1.17$ (Fig.~\ref{fig3}). Although 
chain formation remains the lowest two-particle energy, the large magnetic 
torque at high $\Pi_2$ traps pairs in the rotating dimer metastable minimum 
before they can relax into a chain. As $\Pi_1$ and $\Pi_2$ both increase 
these dimers agglomerate into larger rotating aggregates with elevated $\Phi_V$ 
and $\Phi_A$ but small net displacement; at the highest $\Pi_2$ values 
agglomeration goes to completion, $P$ falls to $\lesssim 0.03$, and all 
vortex order collapses (Figs.~\ref{fig4},\ref{fig5}). An analogous arrest 
occurs in non-active magnetic polymers~\cite{kantorovich2013influence}; here 
self-propulsion delays but cannot prevent the same aggregation pathway.

We have demonstrated that particle elongation and a spatially distributed 
magnetic moment together provide two independent and experimentally accessible 
tuning knobs (charge separation $\Pi_1$ and dipole strength $\Pi_2$) 
that span the full collective phase diagram of magnetic active matter at fixed 
density and activity. The geometric lever arm $\ell$ encoded in $\Pi_1$ 
controls the competition between head-to-tail attraction and antiparallel 
torque, while $\Pi_2$ sets the overall magnetic energy scale; their interplay 
drives transitions among five collective states whose boundaries emerge 
directly from pair contact energies. This establishes a unified and predictive 
energetic framework for collective state selection, and is the magnetic 
analogue of the shape-anisotropy control recently demonstrated for non-magnetic 
active rods~\cite{shelke2026shape}.

The phase diagram recovers and unifies states reported across disparate 
experimental and numerical studies~\cite{kaiser2017, han2020, 
liao2020dynamical, vanesse2023collective, telezki2020simulations} while 
identifying a locked-dimer state (known from equilibrium magnetic 
cylinders~\cite{kantorovich2013influence} but not previously observed in 
active systems) as a distinct non-equilibrium phase driven by kinetic 
trapping. Crucially, all transitions are achieved by adjusting particle 
geometry and magnetization alone, quantities that map directly onto cylindrical 
rare-earth magnets in vibrobot and colloidal 
realizations~\cite{obreque2024dynamics, musacchio2026fluidization, 
sepulveda2021bioinspired}, without requiring changes to density, activity, or 
an external field.

Each phase carries functional implications: the polar flock enables directed 
transport, the vortex-alignment state provides self-sustained rotation for 
microscale mixing and pumping, and the chain and locked-dimer states offer 
programmable self-assembly pathways. Together they position elongated magnetic 
active particles as a versatile platform for on-demand collective 
functionality, complementing recent advances in soft magnetic 
microrobotics~\cite{tierno2021transport, gao2025soft, yan2024programming, 
pramanik2024nature}.

\begin{acknowledgments}
F. G-L acknowledges the support of Fondecyt Regular 1250913.
\end{acknowledgments}



\bibliographystyle{unsrt}

\end{document}